\documentclass[preprintnumbers,amsmath,amssymb]{revtex4}
\usepackage{graphicx}
\newcommand{\tr}{{\rm Tr}}
\begin{document}
\preprint{BNL-NT-04/17}

\title{
Free energy of a static quark anti-quark pair and
the renormalized Polyakov loop in three flavor QCD
}

\medskip

\author{P. Petreczky}
\author{K. Petrov}
\affiliation{Physics Department, Brookhaven National Laboratory, Upton
 NY 11973}

\begin{abstract}
We study the free energy of a static quark anti-quark ($Q \bar Q$)
pair at finite temperature in three flavor
QCD with degenerate quark masses 
using $N_{\tau}=4$ and $6$ lattices with Asqtad 
staggered fermion action. 
The static free energy was calculated for different
values of the quark mass and the entropy contribution at large
distances has been extracted.
We also calculate the renormalized Polyakov loop
following the approach by Kaczmarek et al.
\end{abstract}

\maketitle

\section{ Introduction}

Non-perturbatively the in-medium modification of inter-quark forces,
e.g. screening of fundamental charges at finite temperature, 
is usually studied in terms of the free energy of static (anti)quarks
\cite{attig88,okacz00,ophil02,okacz02,digal03,okacz03,okacz03a}.
More precisely one calculates the difference in the free energy
of the system with static quarks and the same system without quarks
at fixed temperature. The free energy of static quark anti-quark
for different color channels (singlet, octet and averaged)
has been studied in great detail in quenched QCD (pure SU(3) gauge theory)
\cite{attig88,okacz00,okacz02} as well as in 
SU(2) gauge theory \cite{ophil02,digal03}.
In the case of full QCD only the color averaged free energy was extensively
studied; the first results for singlet and octet free
energy for two flavor QCD have appeared only very recently \cite{okacz03a}.
In this paper we are going to study the free energy of static quark anti-quark
pair in 3 flavor QCD using the so-called Asqtad staggered fermion action \cite{asqtad}
with two different lattice spacings (corresponding to 
$N_t=4$ and $6$) at three different quark masses. 

Apart from the in-medium modification of inter-quark forces the 
study of the static free energy is interesting as it can be used
to define the renormalized Polyakov loop. Although the Polyakov
loop is not an order parameter in the presence of dynamical quarks
it shows rapid variation in the transition region and therefore 
is widely used to describe the transition (crossover) in full QCD.
In particular, they are useful for constructing effective mean-field theories 
\cite{dumitru} and studying the interplay between the chiral and deconfining
aspects of the transition \cite{mocsy}. We will study the temperature
dependence of the renormalized Polyakov loop 
which shares most of the properties of the usual Polyakov loop but has
a meaningful continuum limit.

The rest of the paper is organized as follows. In section II we discuss the
lattice setup, parameters of simulations and the zero temperature 
potential which turns out to be crucial for the analysis of the finite
temperature free energies presented in section III. Conclusions are
presented in section IV.

\section{Parameters of simulations and the static potential
at zero temperature}

In this section we are going to describe the lattice parameters for
which our analysis have been performed. 
In our study we use staggered fermions with Asqtad action \cite{asqtad}. 
Our analysis to a large extent is based
on the gauge configurations generated by the MILC collaboration using
the Asqtad action. Therefore we adopt their strategy for fixing the
parameters which is described in Refs. \cite{milc_therm1,milc_therm2,milc_therm3},
namely the strange quark mass was fixed by requiring that $m_{\eta}/m_{\phi}=0.673$, and
the temperature scale (the inverse lattice spacing) was fixed from the scale $r_1$
defined from the zero temperature static potential as 
\begin{equation}
\left(r^2 \frac{d V(r)}{d r}\right)_{r=r_1}=1.
\end{equation}
The RG inspired Ansatz for the gauge coupling dependence of the 
lattice spacing \cite{alton} was used for the scale setting.
We use the most recent value of $r_1$ extrapolated to continuum
and to the physical value of the light quark masses $r_1=0.317$ fm \cite{milc_new}
to convert the lattice units to physical units.
In Table \ref{par_tab} we summarize the lattices, quark masses and the corresponding
temperature range used in our analysis.

\begin{figure}
\includegraphics[width=3in]{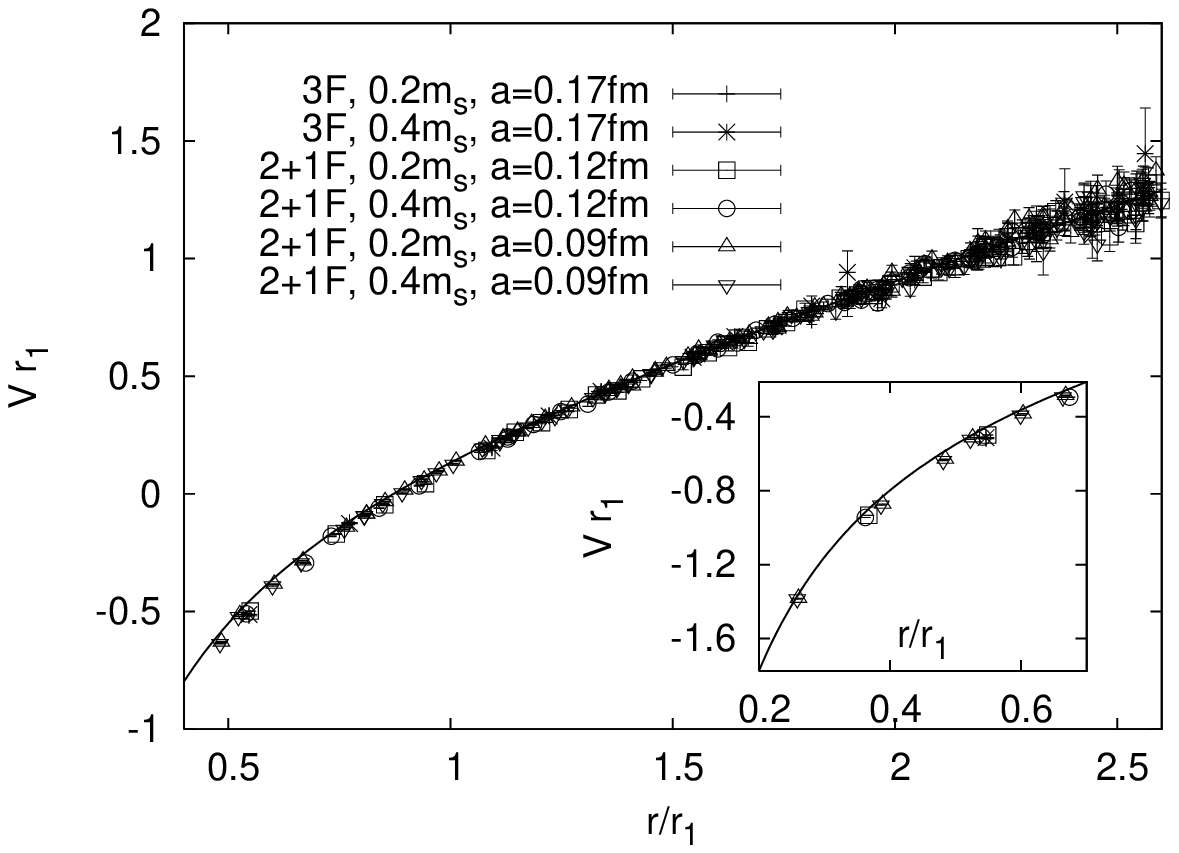}
\includegraphics[width=3in]{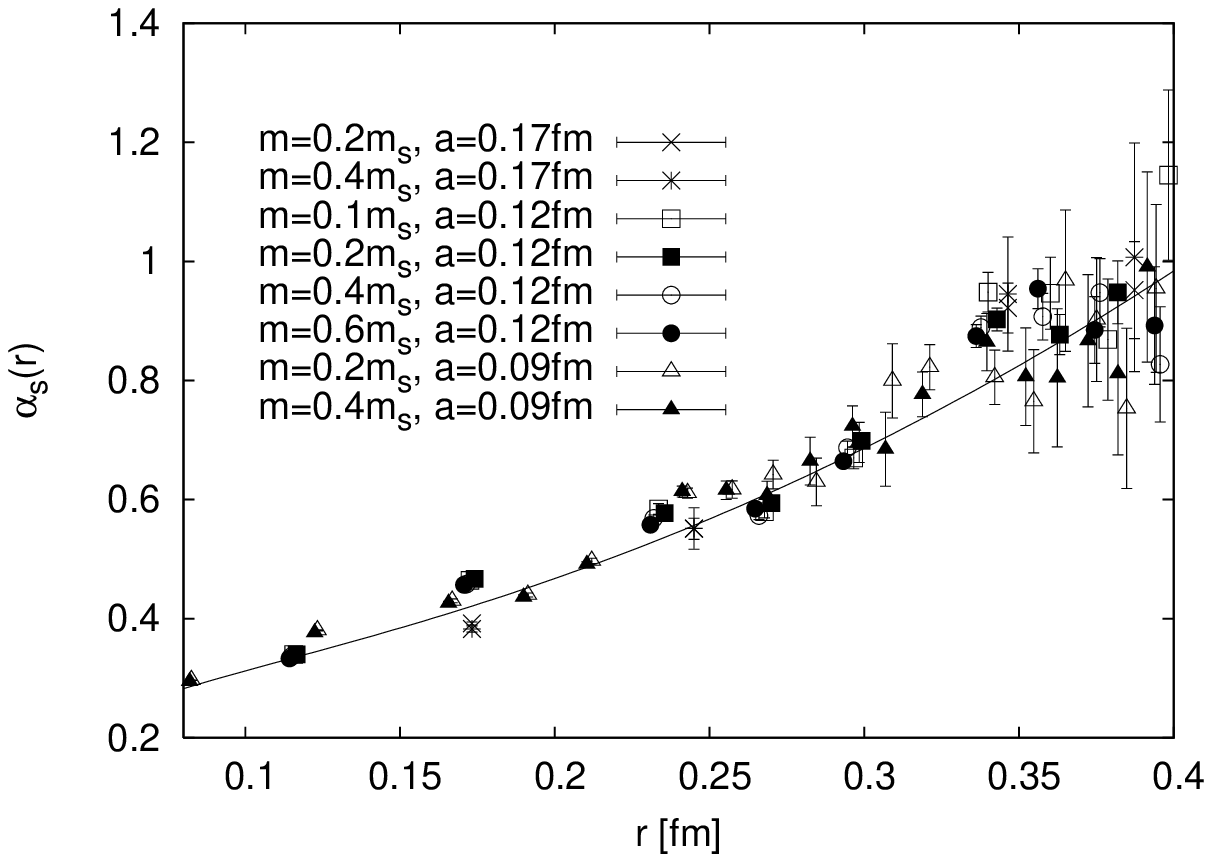}
\caption{The static potential at zero temperature  from the MILC
collaboration versus the parameterization (\ref{pot_ansatz}) (right)
and the effective coupling constant $\alpha_s(r)$ 
as a function of distance (left). Both  cases of three degenerate 
quarks  denoted as 3F as well as one heavy ($m_q=m_s$) and two light denoted as
2+1F are shown.
\label{ztp_fig}}
\end{figure}

The free energy of a static quark anti-quark pair contains a lattice
spacing dependent divergent piece and thus needs to be renormalized.
This can be done by normalizing it to the zero temperature potential at
short distances where the temperature dependence of the free energy 
can be neglected \cite{okacz02}. The static quark potential has been
studied by the MILC collaboration at three different lattice spacings
and various quark masses \cite{milc_therm3,milc_new,milc_pot1,priv}. 
The zero temperature potential  is also defined
only up to some constant which needs to be fixed. As temperature 
is varied by varying the lattice spacing we need to specify the form of
the zero temperature potential which is valid for the whole range of lattice
spacing relevant in this study, i.e. $0.09fm < a < 0.3 fm$.  
We choose the following form of the zero temperature
potential
\begin{equation}
r_1 V(x)= -\frac{0.44}{x}+0.56 \cdot x+\frac{0.0125}{x^2},~ x=r/r_1
\label{pot_ansatz}
\end{equation}
In Fig. \ref{ztp_fig} we show this parametrization of the potential against
lattice data at three different lattice spacing, and different quark masses.
Although in the present paper we discuss the case of three degenerate
flavors, we show the zero temperature potential also for $2+1$ flavor
case as here more detailed data are available. Because of the very weak quark mass
dependence the case of $2+1$ flavors can be used as a good reference.
As one can see this parameterization gives a fare description
of the data for all lattice spacings and quark masses \cite{priv}.
The last term in Eq. (\ref{pot_ansatz}) mimics the effect of the 
running coupling. In Fig. \ref{ztp_fig} we also show the effective
coupling constant $\alpha_s(r)$ defined as 
\begin{equation}
\alpha_s(r)=\frac{3}{4} r^2 \frac{d V(r)}{d r}.
\label{effalpha}
\end{equation}
One can see again that Ansatz (\ref{pot_ansatz}) gives a good description
of the lattice data.
We note that in order to reduce the effects of the breaking of the rotational
symmetry in the heavy quark potential following
Refs. \cite{okacz02,necco} the separation $r$ between the static
charges was redefined as $r \equiv r_I= (4 \pi C_L(r))^{-1}$, where
$C_L(r)$ is the lattice Coulomb potential for the Luescher-Weisz action
\cite{sym}. In what follows $r$ will always refer to the modified 
separation defined above.

\begin{table}
\begin{tabular}{|l|l|l|}
\hline
lattice & $m_{light}$  & $T$ \\
\hline
$12^3 \times 4$ & $0.2m_s$, $0.4m_s$, $0.6m_s$ & $135-393$ MeV\\
\hline
$8^3 \times 4$  & $0.2m_s$, $0.4m_s$, $0.6m_s$ & $135-412$ MeV\\
\hline
$12^3 \times 6$ & $0.2m_s$, $0.4m_s$, $0.6m_s$ & $145-310$MeV\\
\hline
\end{tabular}
\caption{The parameters of simulations, $m_s$ denotes the strange quark mass.
\label{par_tab}}
\end{table}

\section{The free energy of static quark anti-quark}

Following Refs. \cite{mclerran,nadkarni} the free energy of static 
quark-antiquark ( $Q \bar Q$) pair, i.e. the free energy difference of the system with and 
without static sources at fixed temperature $T$ in the color singlet and octet channels is
defined as
\begin{eqnarray}
&&
\exp(-F_1(r,T)/T+C)=\frac{1}{3} \tr \langle W(\vec{r}) W^{\dagger}(0) \rangle\\
&&
\exp(-F_8(r,T)/T+C)=\frac{1}{8} \langle \tr W(\vec{r}) \tr W^{\dagger}(0) \rangle-
\frac{1}{24} \tr \langle W(\vec{r}) W^{\dagger}(0) \rangle
\end{eqnarray}
Here $W(\vec{x})=\prod_{\tau=0}^{N_{\tau}-1} U_0(\tau,\vec{x})$ is the temporal 
Wilson line and $L(\vec{x})=\tr W(\vec{x})$ is known as the Polyakov loop.
As $W(\vec{x})$ is not gauge invariant one needs to fix a gauge in order
to calculate $F_1$ and $F_8$.  In this study we use the Coulomb gauge as for
this gauge a transfer matrix  can be constructed and in the zero temperature
limit the usual static potential will be recovered. Alternatively one can 
replace the Wilson line $W(\vec{x})$ by a dressed gauge invariant Wilson line
$\tilde W(\vec{x})$ using the eigenvectors of covariant spatial Laplacian
\cite{ophil02}. The dressed Wilson line, however,  is a non-local operator. 
Both definitions turned out to equivalent \cite{ophil02}
(at least numerically).
One can also consider the color averaged free energy defined as
\begin{equation}
\exp(-F_{av}(r,T)/T+C)=\frac{1}{9} \langle \tr W(\vec{r}) \tr W^{\dagger}(0) \rangle
\equiv \frac{1}{9} \langle L(\vec{r}) L^{\dagger}(0) \rangle
\end{equation}
which can be written as a thermal average of the singlet and octet free energies
\begin{equation}
\exp(-F_{av}(r,T)/T)=\frac{1}{9} \exp(-F_{1}(r,T)/T)+\frac{8}{9} \exp(-F_{8}(r,T)/T).
\end{equation}
This quantity is expressed in terms of local explicitly gauge invariant operators
and for this reason was extensively studied in the past also in full QCD
\cite{detar99,karsch01}. The normalization constant $C$ needs to be fixed
using some physical normalization condition. As we expect that at very short distances
the free energy of static quark anti-quark pair is only given by their
interaction energy in the vacuum we fix $C$ by requiring that the singlet free
energy coincides with the zero temperature potential given by Eq. (\ref{pot_ansatz}).

\begin{figure}
\includegraphics[width=3.3in]{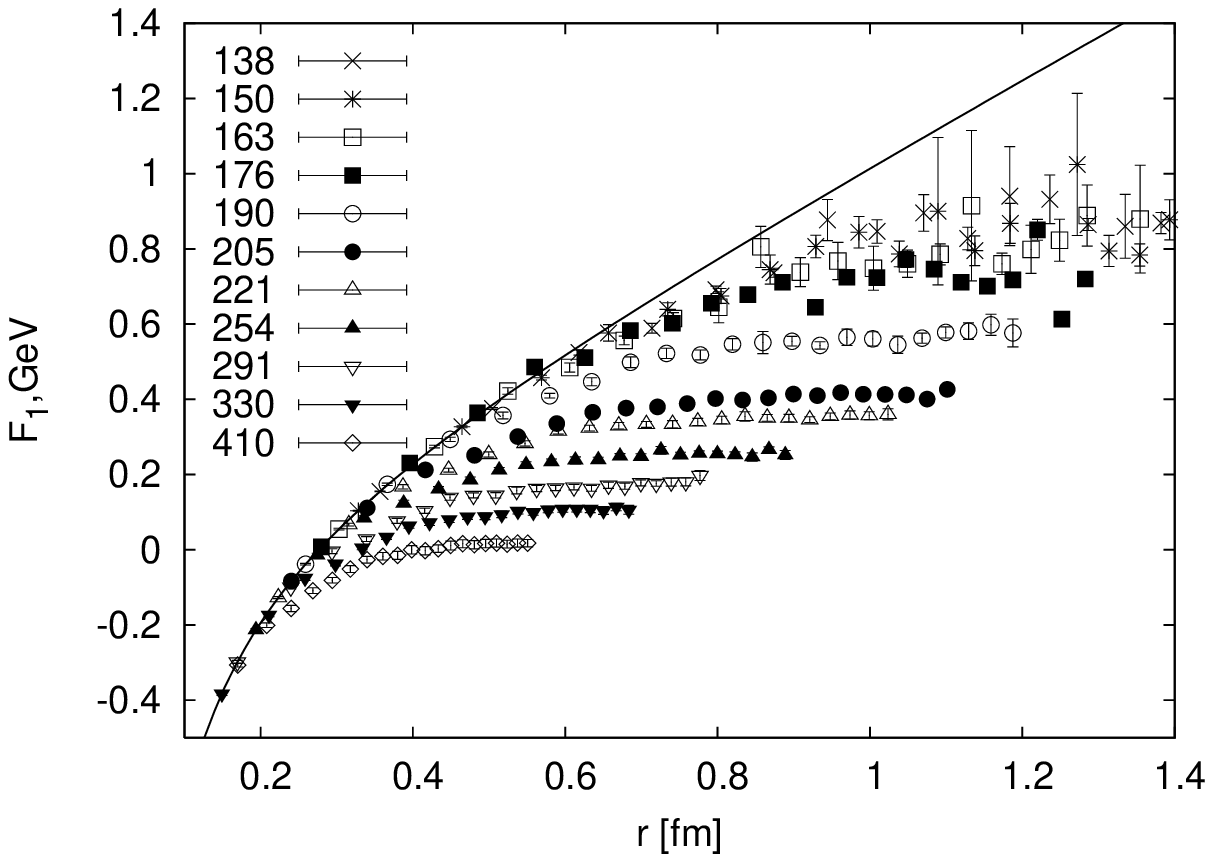}
\includegraphics[width=3.3in]{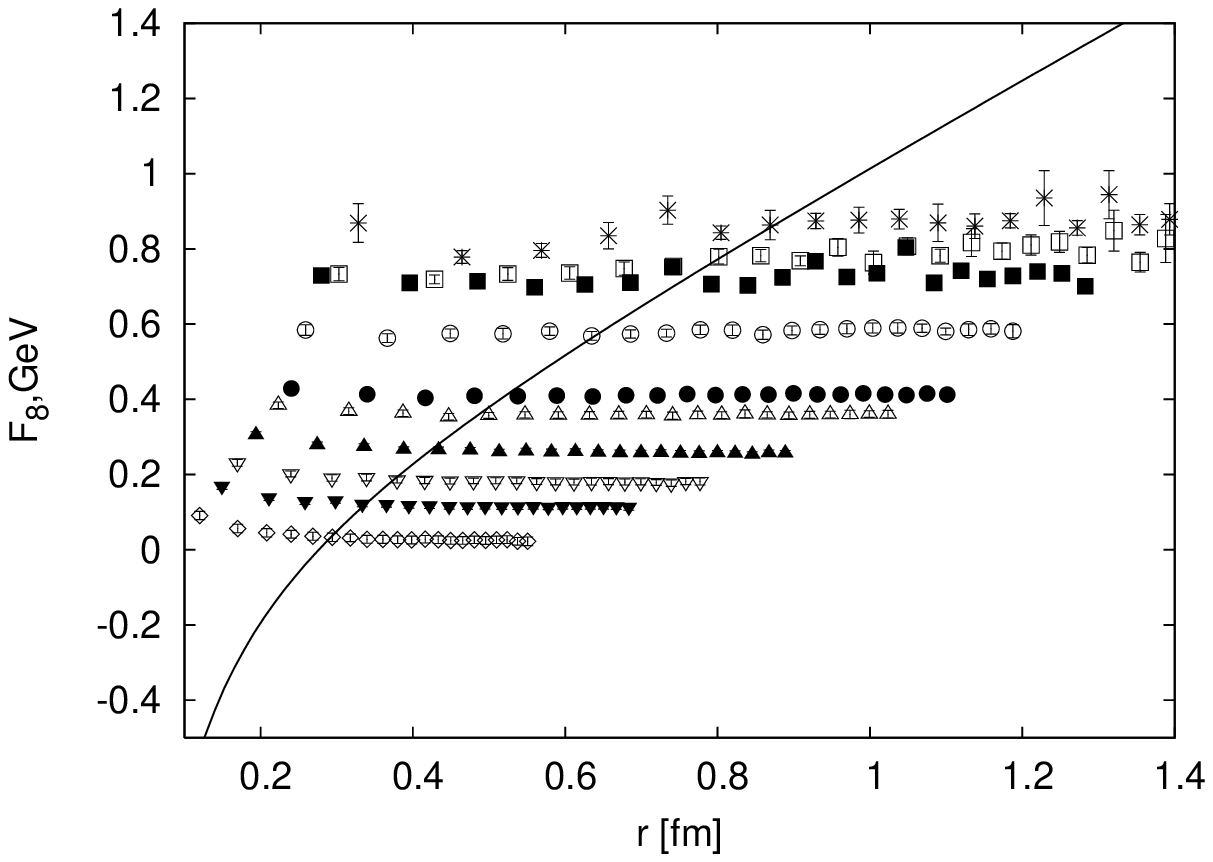}
\caption{The singlet (left) and octet (right) free energy of static $Q\bar Q$
pair calculated on $12^3 \times 4$ lattice at $m_q=0.4m_s$ at different
temperature in MeV. The solid line
is the parameterization of the zero temperature potential by Eq. \ref{pot_ansatz}.
\label{fig_f04ms_124}}
\end{figure}

We start the discussion of our numerical results with the case of the
quark of mass $0.4m_s$ on the $12^3 \times 4$ lattice
\footnote{
Most of the configurations used for our measurements were provided to 
us by the MILC collaboration.}.
The corresponding numerical results
for the singlet and octet free energy are shown in Fig. \ref{fig_f04ms_124}.
The singlet free energy approaches a finite value at large distances which is
usually interpreted as string breaking at low temperature and screening at
high ones. Note that the distance where the free energy effectively flattens
is temperature dependent; it becomes smaller at higher temperatures. At small
distances the singlet free energy is temperature independent and coincides with the
zero temperature potential. This is intuitively expected as at small distances medium effects
are not important. Similar observation has been made for quenched QCD 
\cite{okacz02,okacz03,okacz03a}. The octet free energy shows much stronger
temperature dependence. At short distances it is expected to have a repulsive tail,
which is clearly visible at high temperatures.  At low temperatures the presence of such a 
repulsive tail is less obvious. The reason for this is the following. At low temperatures
no data is available at really short distances ($<0.2fm$)
and the (large distance ) asymptotic value of the octet free energy
is large compared to the value of the repulsive tail (which  naively is
$\alpha_s/(6r)$). Also the statistical accuracy is lowest 
at low temperatures. As the temperature increases both the value of the free energy 
at large distances  is smaller and the statistical accuracy is higher; in addition more
data at shorter distances become accessible. 
The numerical results for other values of the quark masses are similar.

\begin{figure}
\includegraphics[width=2in]{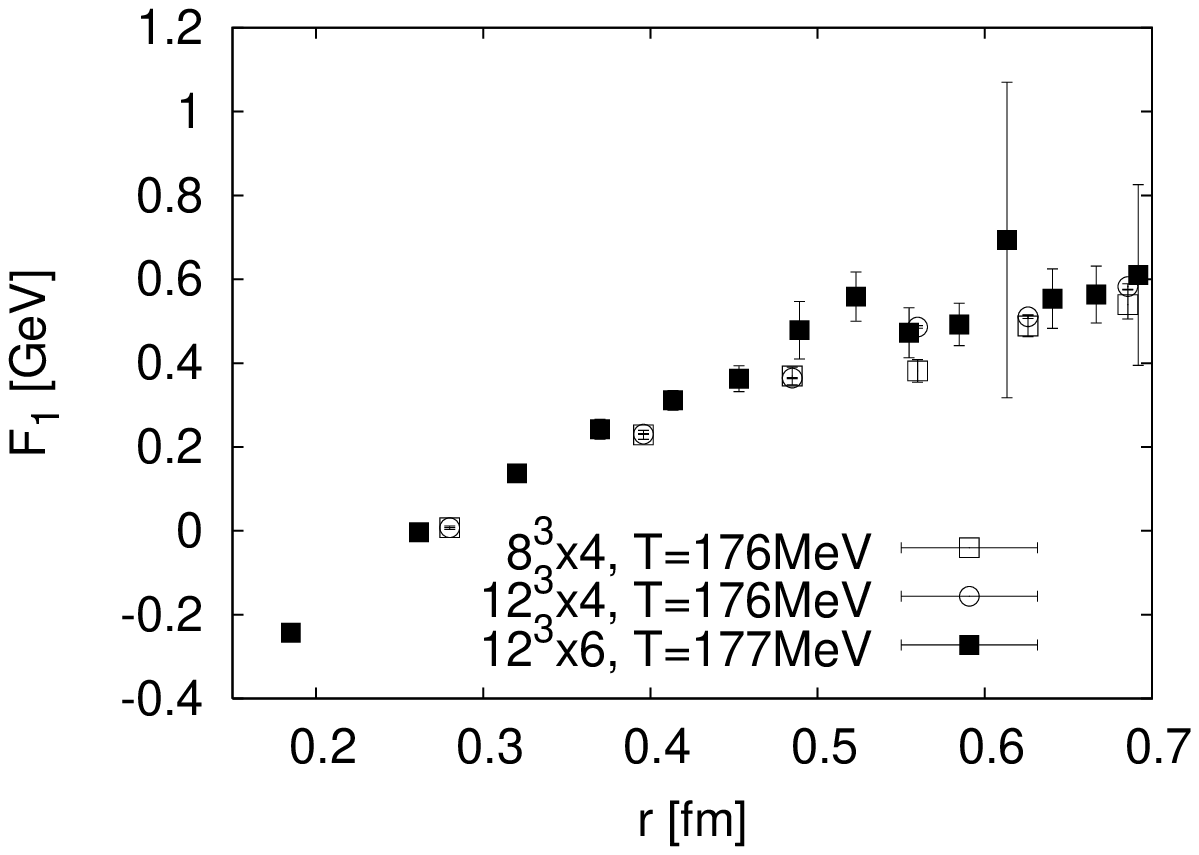} \includegraphics[width=2in]{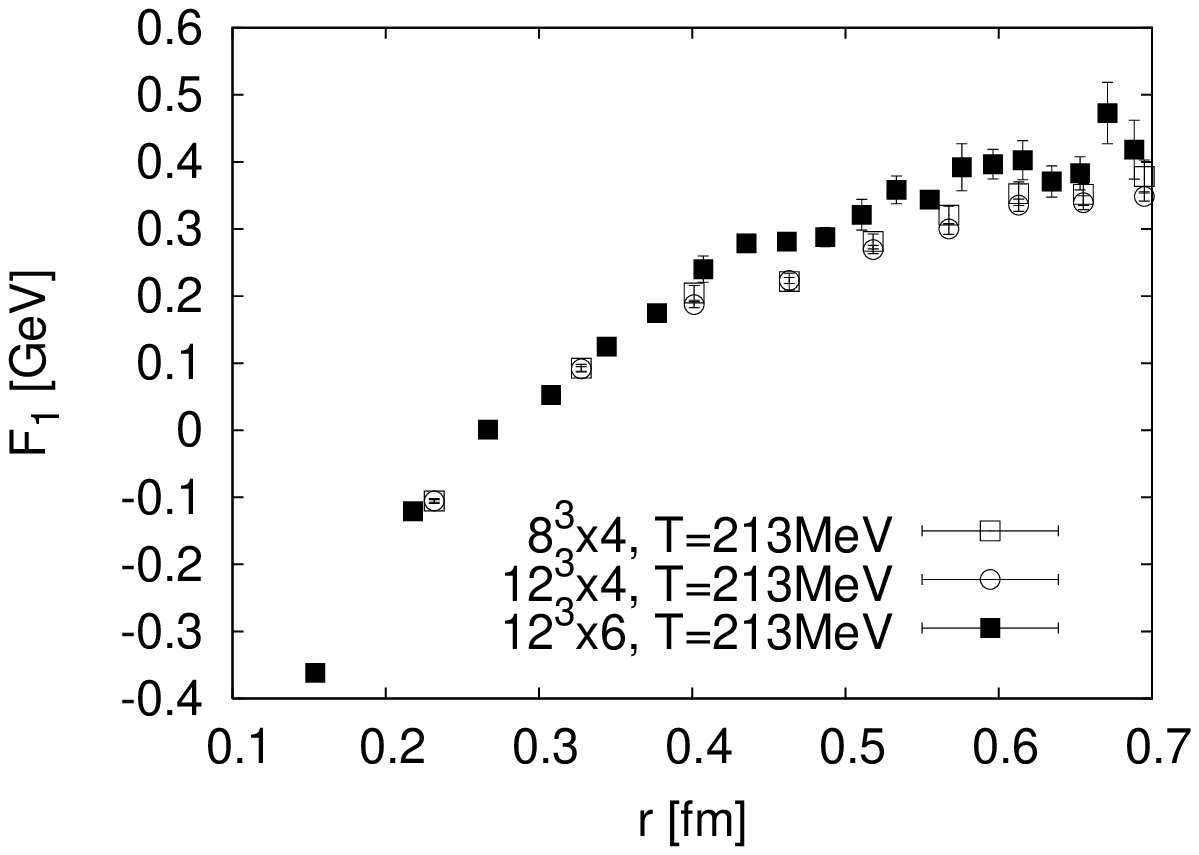} 
\includegraphics[width=2in]{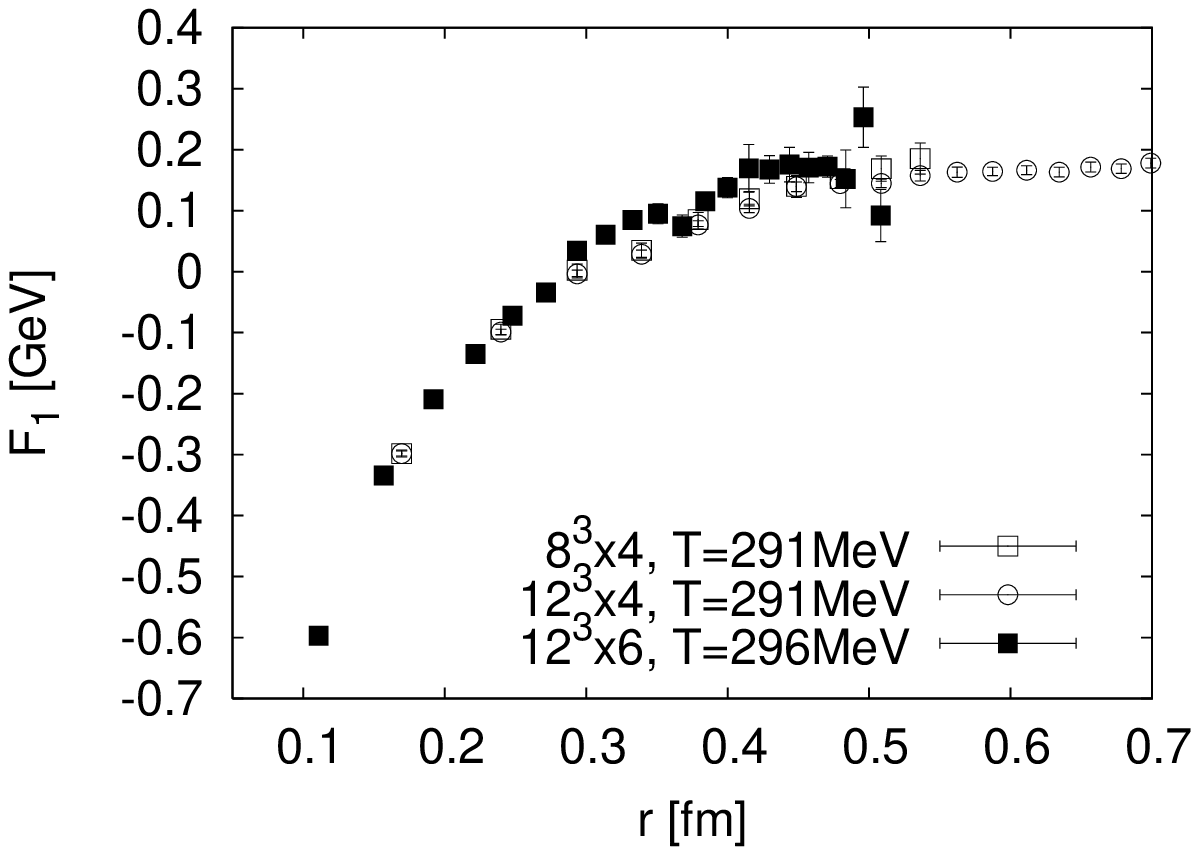}
\caption{
The singlet free energy calculated on $8^3 \times 4$, $12^3 \times 4$ and
$12^3 \times 6$ lattices at three different temperatures.
\label{fig_fscaling}
}
\end{figure}

As $N_{\tau}=4$ lattices correspond to a quite large lattice spacing, one may worry
about lattice artifacts in the static free energies. Therefore
we have compared the singlet free energies calculated from $N_{\tau}=4$
and $N_{\tau}=6$ lattices at approximately same temperatures as shown in
Fig. \ref{fig_fscaling}. As one can see, there is no sizeable lattice
spacing dependence even at the lowest temperatures where the lattice
spacing is the largest. Thus calculations of the quark anti-quark free
energy on $N_{\tau}=4$ lattices are justified.

\begin{figure}
\includegraphics[width=3in]{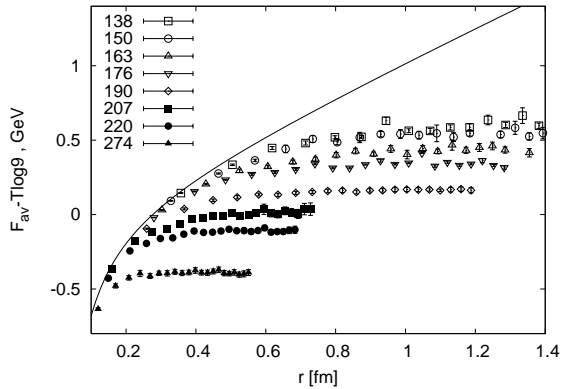}
\caption{The color averaged free energy at different temperatures calculated
on $N_{\tau}=4$ (open symbols) and  $N_{\tau}=6$ (filled symbols) lattices.
The solid line is the zero temperature potential.\label{fig_fa}}
\end{figure}

We also investigate the temperature dependence of the color averaged
free energy. In general due to the presence of color octet contribution
the temperature dependence of the color averaged free energy 
is stronger than that of the singlet free energy \cite{karsch01}.
At small distances, however, it is expected to be dominated by color singlet contribution
and the approximate relation $F_{av}(r,T) \simeq F_1(r,T)+T \ln 9$ should hold \cite{okacz02}.
Therefore in Fig. \ref{fig_fa} we show $F_{av}(r,T)-T \ln 9$ at different temperatures
including results both from $N_{\tau}=4$ and $6$ lattices. As one can see from the figure
for temperatures $T<274MeV$ $F_{av}-T \ln 9$ coincides with the zero temperature
potential at the shortest distance available in this study. As at the shortest distance the
zero temperature potential coincides with the singlet free energy (due to normalization)
this in turn implies that the relation $F_{av}=F_1+ T \ln 9$ holds at this distance.

As expected, the temperature dependence of the color averaged free energy is stronger
than of the color singlet one.

From Figs.\ref{fig_f04ms_124},\ref{fig_fa} we see that the free energy 
for all color channels reach a constant
value at large distances $F_{\infty}^i(T)=\lim_{r \rightarrow \infty} F_i(r,T),~~i=1,8,av$.
In the infinite volume limit we expect that
$F_{\infty}^1(T)=F_{\infty}^8(T)=F_{\infty}^{av}(T)$ as at very large distances the 
free energy of a static quark and anti-quark should not depend on their 
relative color orientation. This is in fact confirmed in the deconfined phase of
pure gauge theory \cite{okacz02,okacz03,digal03}. On our small lattices, $8^3 \times 4$,
$12^3 \times 6$ we see small differences between $F_{\infty}^1$ and 
$F_{\infty}^{8,av}$ which vanish within statistical errors when we go to $12^3 \times 4$ 
lattice. In practice we determine $F_{\infty}^i(T)$ by fitting the 
corresponding free energies by a constant at large distances.
The lower limit of the fit range was always chosen such that the 
determined value of $F_{\infty}^i$ does not depend on it within our statistical accuracy.

The Polyakov loop $L(x)=\tr W(x)$ is the order parameter
for deconfining transition in pure gauge theories. In full
QCD, where dynamical quarks are present, it is no longer an
order parameter as it has non-zero value in low 
temperature phase. Nevertheless, it is used to study
the deconfining aspects of the transition in full QCD as
well in effective models \cite{dumitru,mocsy}. 
The correlator of Polyakov loops corresponding to the color averaged free
energy satisfies the cluster decomposition
\begin{equation}
\langle L(r) L^{\dagger}(0) \rangle|_{r \rightarrow \infty}= |\langle L (0)\rangle|^2.
\end{equation}
Therefore following Ref. \cite{okacz02}, we can define the renormalized
Polyakov loop as 
\begin{equation}
L_{ren}=\exp(-\frac{F_{\infty}^{av}(T)}{2 T}).
\end{equation}
The numerical results for $L_{ren}(T)$ will be discussed in detail at
the end of this section.
In three flavor QCD for the quark masses used in this study
there is no phase transition but only a crossover 
\cite{milc_therm1,milc_therm2,milc_therm3,karsch_endpoint}.
Nevertheless, both the chiral condensate and the unrenormalized Polyakov
loop show a rapid change at approximately the same temperature
\cite{milc_therm1,milc_therm2,milc_therm3,karsch01} referred to
as the transition temperature $T_c$. Therefore we can define 
the transition temperature  as the temperature where 
$\partial L_{ren}(T)/\partial T$ has a maximum, or equivalently 
$\partial F_{ren}(T)/\partial T$ has a minimum. We have determined 
$T_c$ in this way and have found it to be consistent with
the value of $T_c$ defined from the peak of the chiral
susceptibility \cite{milc_therm1,milc_therm2,milc_therm3}. In the
following discussion we will find instructive to plot
different quantities as function of $T/T_c$, especially when comparing
with calculations done in SU(3) gauge theory. 

\begin{figure}
\includegraphics[width=3in]{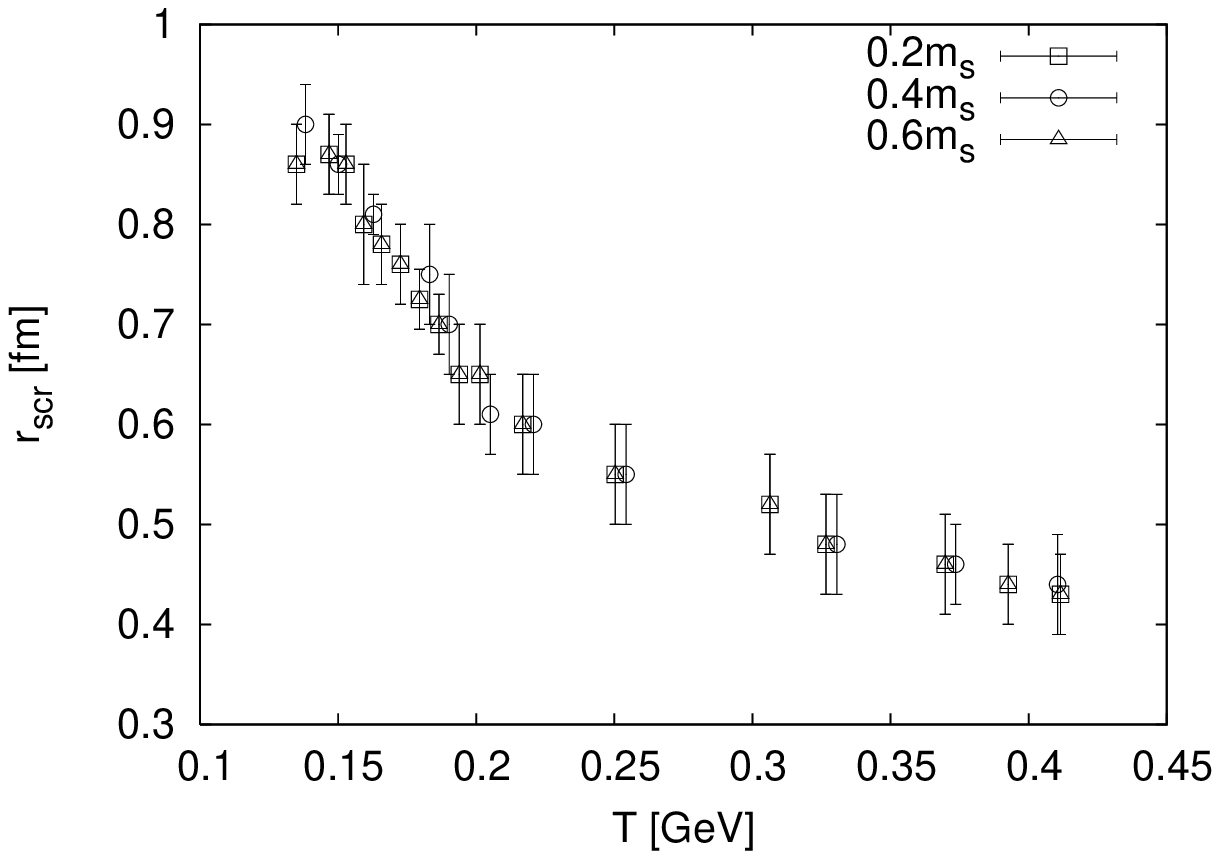}
\includegraphics[width=3in]{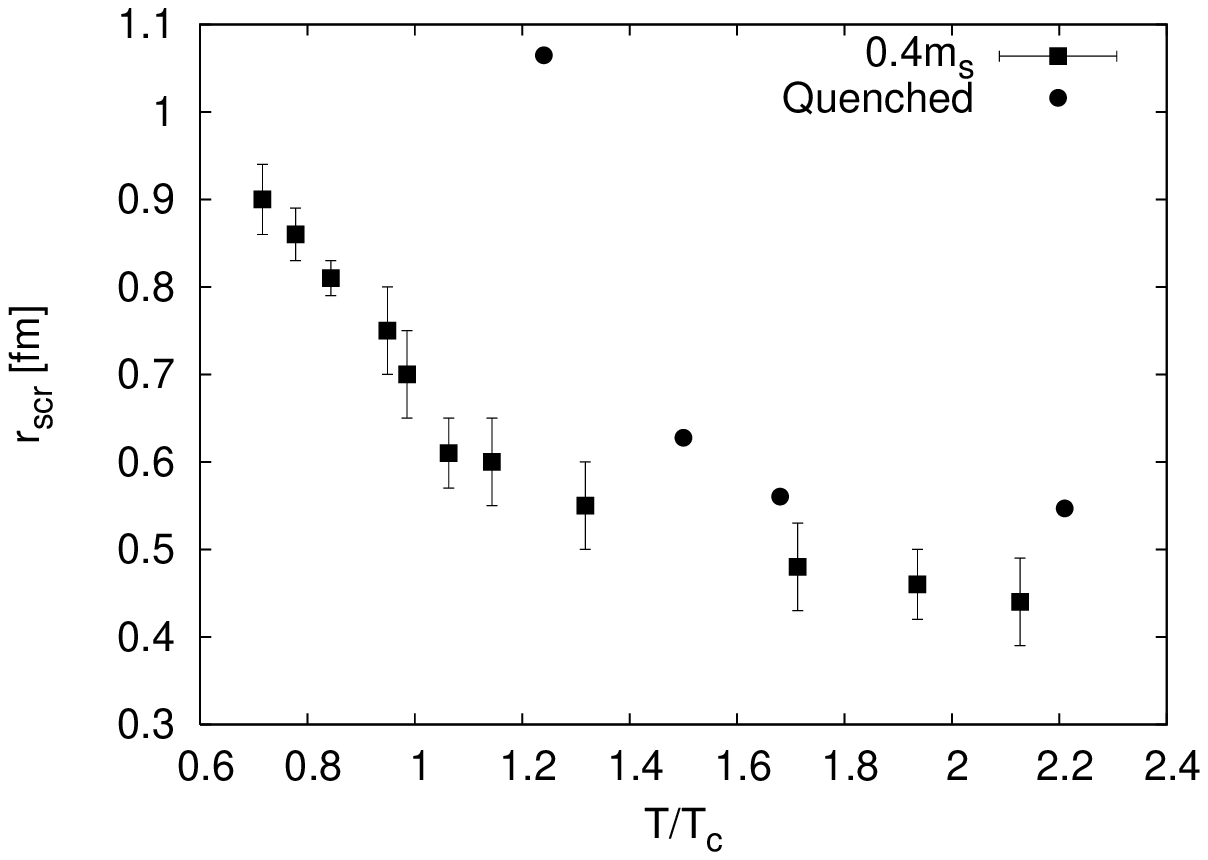}
\caption{The effective screening radius from the singlet free energy as
a function of the temperature for three different quark masses (left)
and as function of $T/T_c$ for $0.4m_s$ and quenched QCD (right)
\label{fig_rscr}
}
\end{figure}

To
characterize the range of interaction in the medium it is convenient to 
introduce the effective screening radius $r_{scr}$. It is defined as the distance
at which the singlet free energy is only $10\%$ below its asymptotic value
$F_1(r=r_{scr},T)=0.9 F_{\infty}^1(T)$. Here $F_{\infty}^1(T)$ is the asymptotic value
of the singlet free energy at infinite separation. In Fig. \ref{fig_rscr}. we show the
values of $r_{scr}$ for three different quark masses and $12^3\times 4$ lattices.
Certainly as $F_1(r,T)$ has statistical errors, it is difficult to determine 
exactly at which distance $r$ the equation $F_1(r=r_{scr},T)=0.9 F_{\infty}^1(T)$ holds.
We have tried to estimate this uncertainties in the values of $r_{scr}$ 
and show them in Fig. \ref{fig_rscr} as errors bars.
At small temperatures the value of the screening radius is about $0.9fm$ and is
temperature independent. As we increase the temperature $r_{scr}$ decreases 
reaching the value of $0.5fm$ at the highest temperature. Note that 
the temperature dependence of $r_{scr}$ is roughly the same for all quark masses.
In Fig. \ref{fig_rscr}  we also show the effective screening radius for $0.4m_s$ and
quenched QCD estimated from the singlet free energies of Ref. \cite{okacz02}
as function of $T/T_c$. Close to $T_c$ the screening
radius in quenched QCD is considerably higher than in three flavor
QCD but at higher temperatures they are comparable.

\begin{figure}
\includegraphics[width=3in]{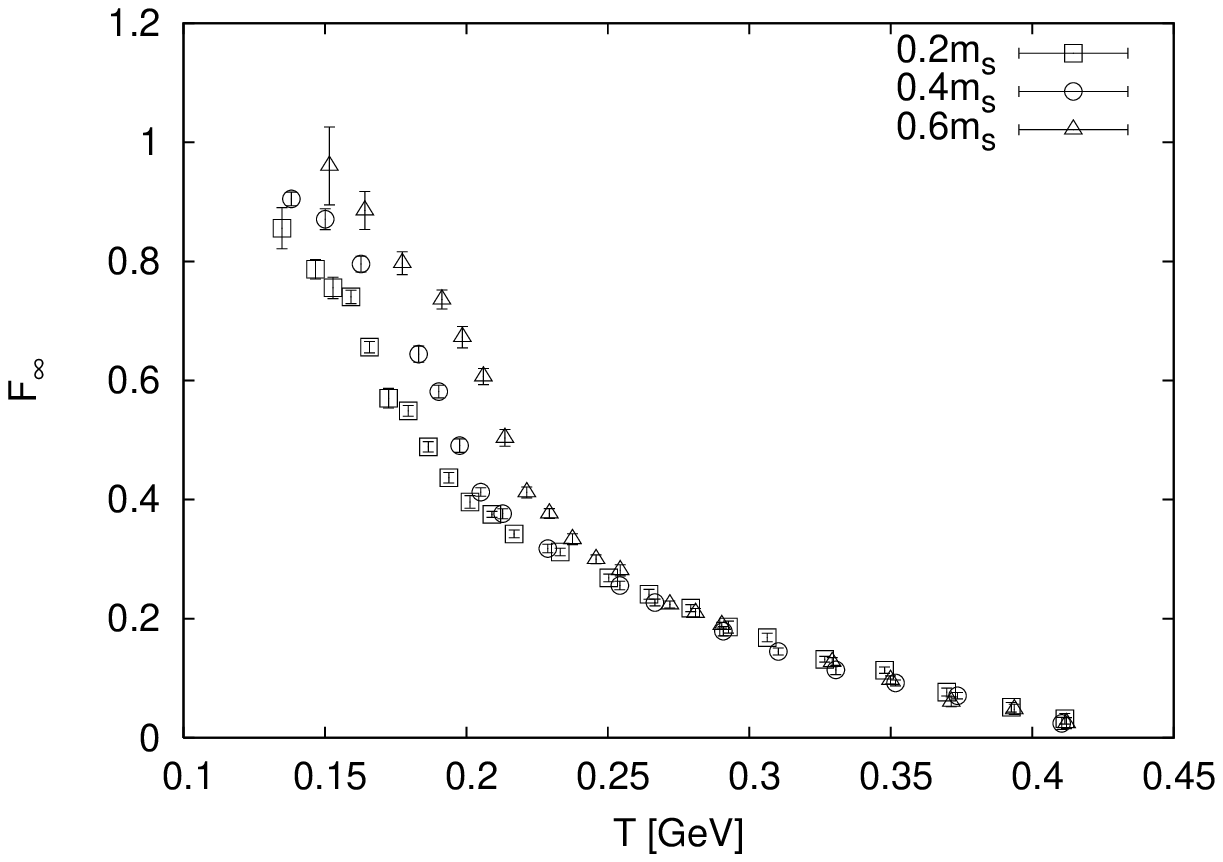}
\includegraphics[width=3in]{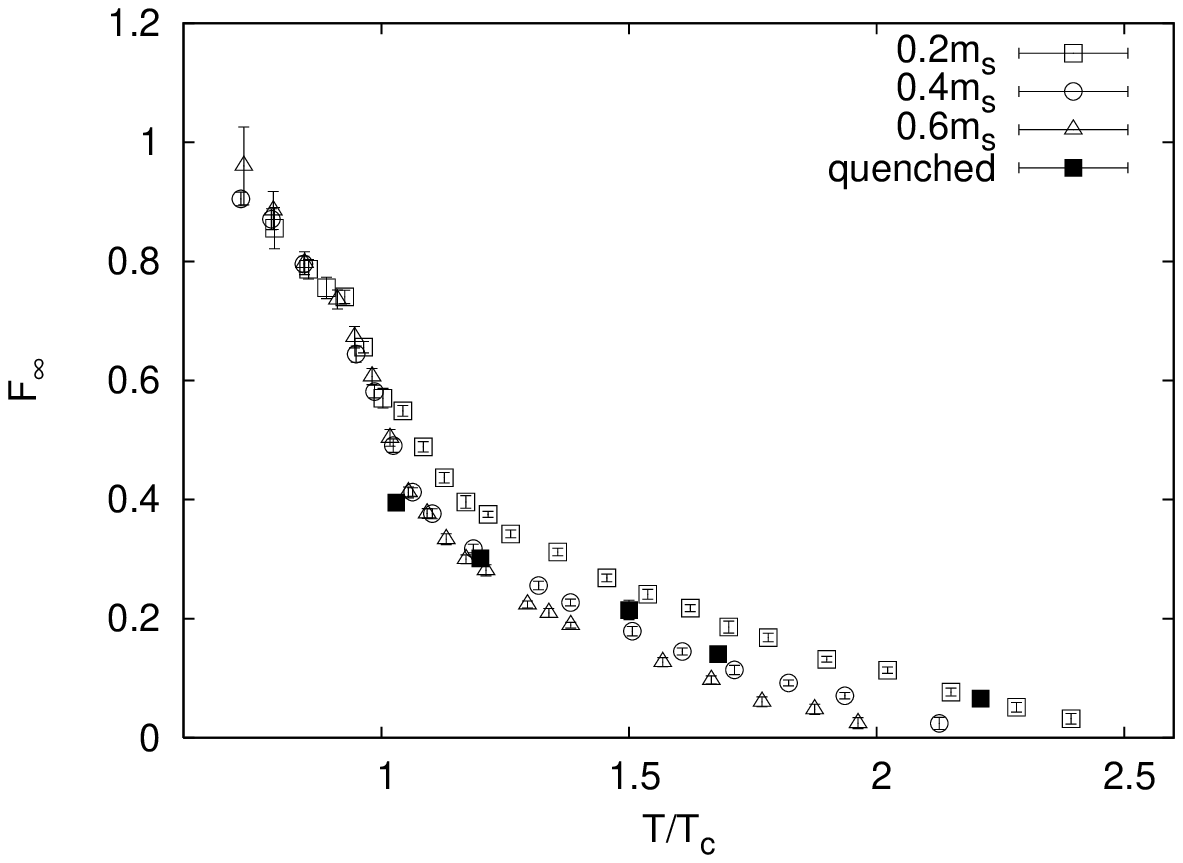}
\caption{The asymptotic value of the free energy at large
distance for different quark masses as function of the
temperature in physical units(left)  and of $T/T_c$ (right).
On the right plot we also show the value of $F_{\infty}$ from
pure gauge theory \cite{okacz02,okacz03}.
\label{fig_finf}}
\end{figure}

Now we are going to investigate the temperature and quark mass dependence
of the asymptotic value of the free energy and identify the entropy contribution to it.
In the following discussion we will
take the asymptotic value of the free energy in the infinite 
volume limit  to be given by the color averaged value 
$F_{\infty}(T) \equiv F_{\infty}^{av}(T)$ and also skip the index $i$ in the
following discussion. 
The numerical results for $F_{\infty}(T)$ at three different quark masses
are shown in Fig. \ref{fig_finf}. One immediately notices that the quark 
mass dependence may vanish at small temperatures ($T<150MeV$) and definitely
negligible at high temperatures ($T>250MeV$) while large mass dependence is observed in the
transition region. One may wonder to which extent this mass dependence is
due to the mass dependence of the transition temperature $T_c$. 
Therefore in Fig. \ref{fig_finf}
we also show the data for $F_{\infty}(T)$ versus $T/T_c$ which shows that up
to $T/T_c=1$ there is no mass dependence. This is in accordance with finding
of Ref. \cite{karsch01} where for quark masses below the strange quark mass 
the mass dependence of $F_{\infty}$ is quite small for $T/T_c$ close to one.
Thus for temperatures below $T_c$ the mass dependence of $F_{\infty}(T)$
can be understood in terms of mass dependence of $T_c$. However,
as one can also see from Fig. \ref{fig_finf}, for $T/T_c>1$ 
substantial mass dependence is seen.

\begin{figure}
\includegraphics[width=3in]{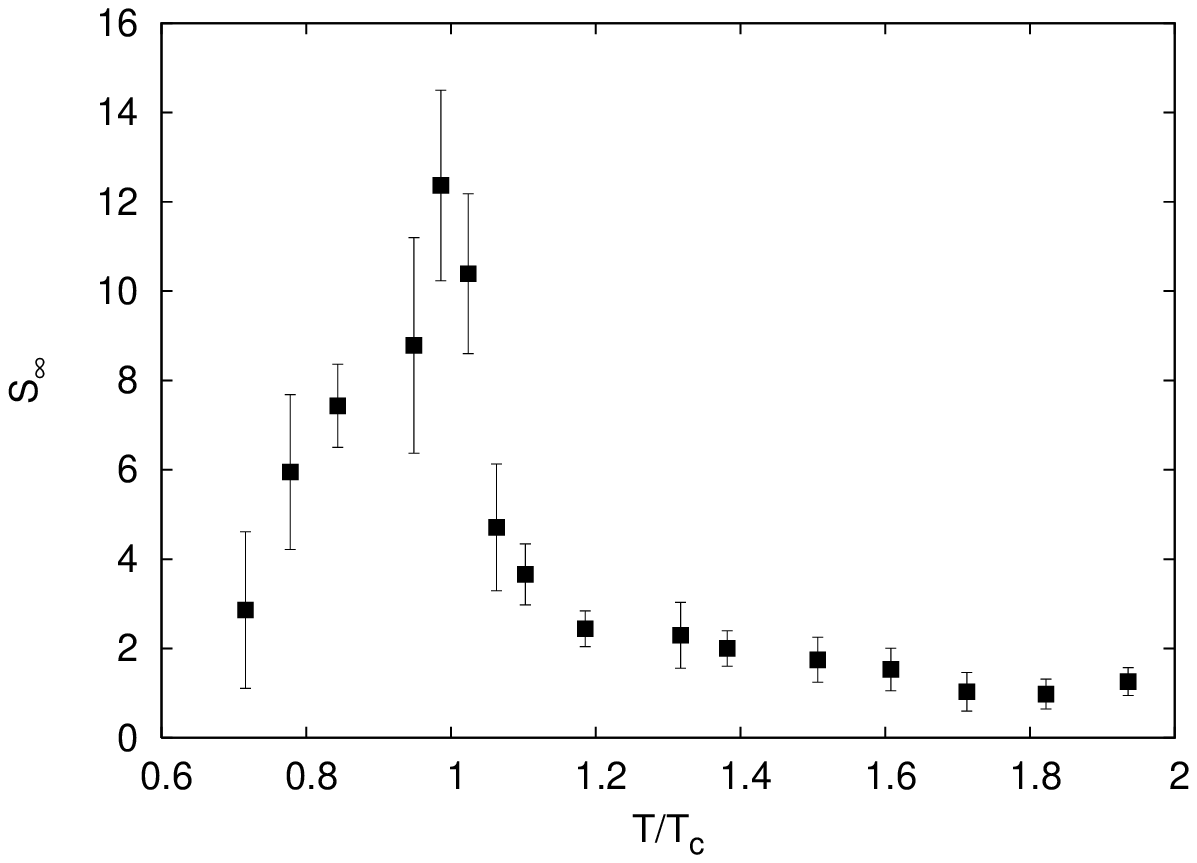}
\includegraphics[width=3in]{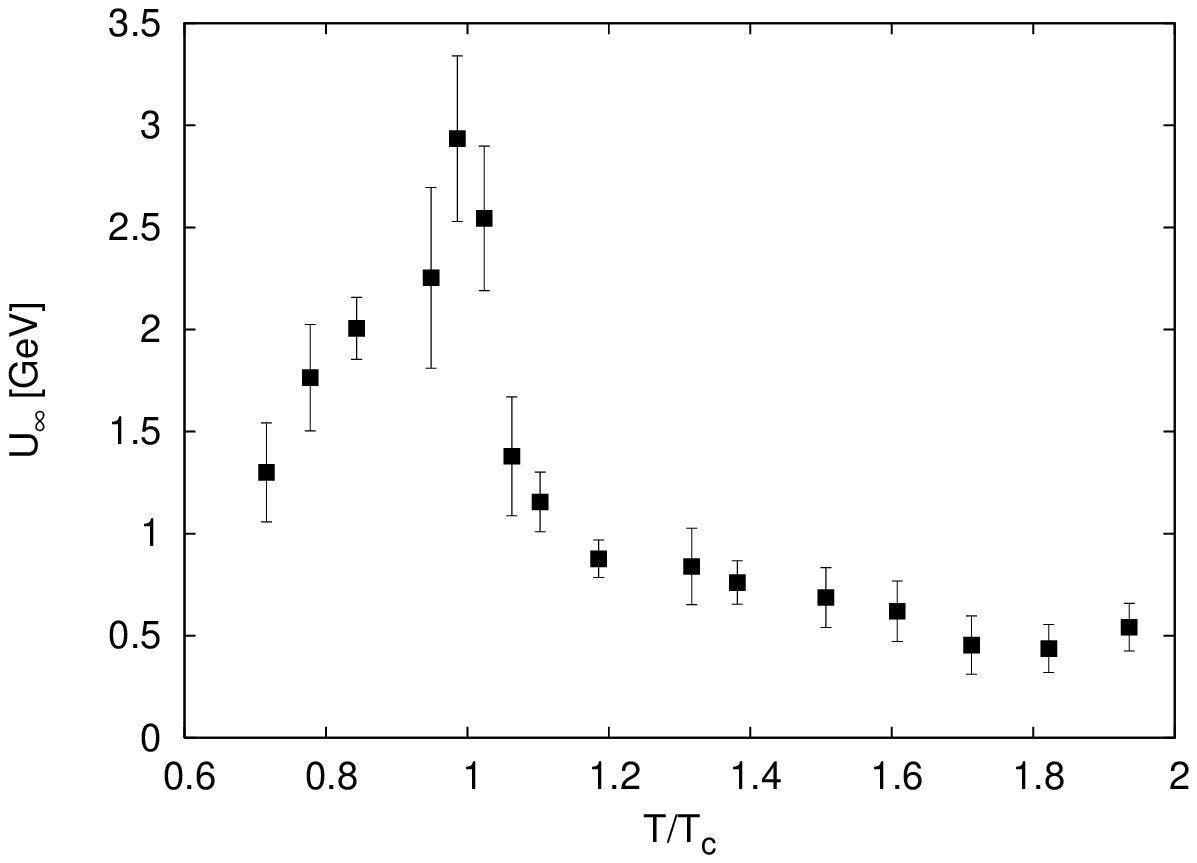}
\caption{The asymptotic values of the entropy and the internal energy of static
$Q \bar Q$ pair at large distances as function of the temperature. \label{fig_thermo}}
\end{figure}

In the case of very small temperatures we expect 
$F_{\infty}(T)$ to be temperature independent and related to 
twice the binding energy of a heavy-light ($D-$ or $B-$) meson
$2 E_{bin}=2 M_{D,B}-2m_{c,b}$. More precisely, it should be
the binding energy of a static-light system as found 
in the 3d SU(2) Higgs model which is quite similar to QCD \cite{ophil98}.
Because of the heavy quark symmetry we expect  the binding
energy of a heavy-light meson to be the same as of the static-light system.
Based on this observation it has been argued in Ref. \cite{digal01} that
the decrease of $F_{\infty}(T)$ with the temperature close to $T_c$ implies
the decrease of the $M_{D,B}$ leading to quarkonium suppression.
However, one should keep in mind that $F_{\infty}(T)$ also contains an
entropy contribution. The entropy due to the presence
of a static $Q \bar Q$ pair can be calculated as
\begin{equation}
S_{\infty}(T)=-\frac{\partial F_{\infty}(T)}{\partial T}.
\label{sinf_eq}
\end{equation}
We also can calculate the energy induced by a static $Q \bar Q$ pair
$U_{\infty}=F_{\infty}+T S_{\infty}$.
Numerically the derivative with respect to the temperature in 
Eq. (\ref{sinf_eq}) was estimated using forward differences.
In Fig. \ref{fig_thermo} we show the entropy $S_{\infty}$ and
energy $U_{\infty}$ as functions of temperature.  Both
the entropy and the energy show a strong  increase near $T_c$ and
with our definition of $T_c$  peak exactly at $T_c$. This large increase in entropy and
energy is probably due to many-body effects and makes the
interpretation of $U_{\infty}$ as the binding energy of
heavy-light meson not very plausible.

\begin{figure}
\includegraphics[width=3in]{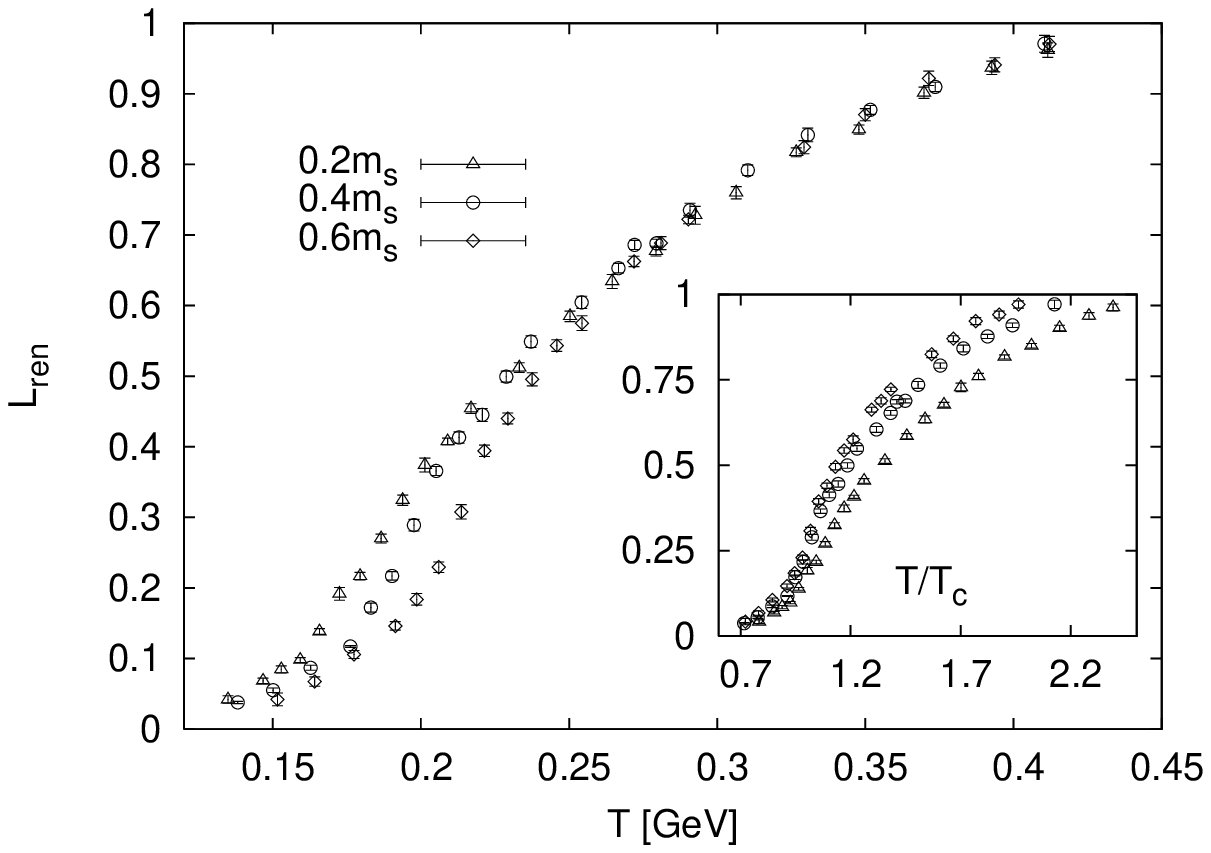}
\includegraphics[width=3in]{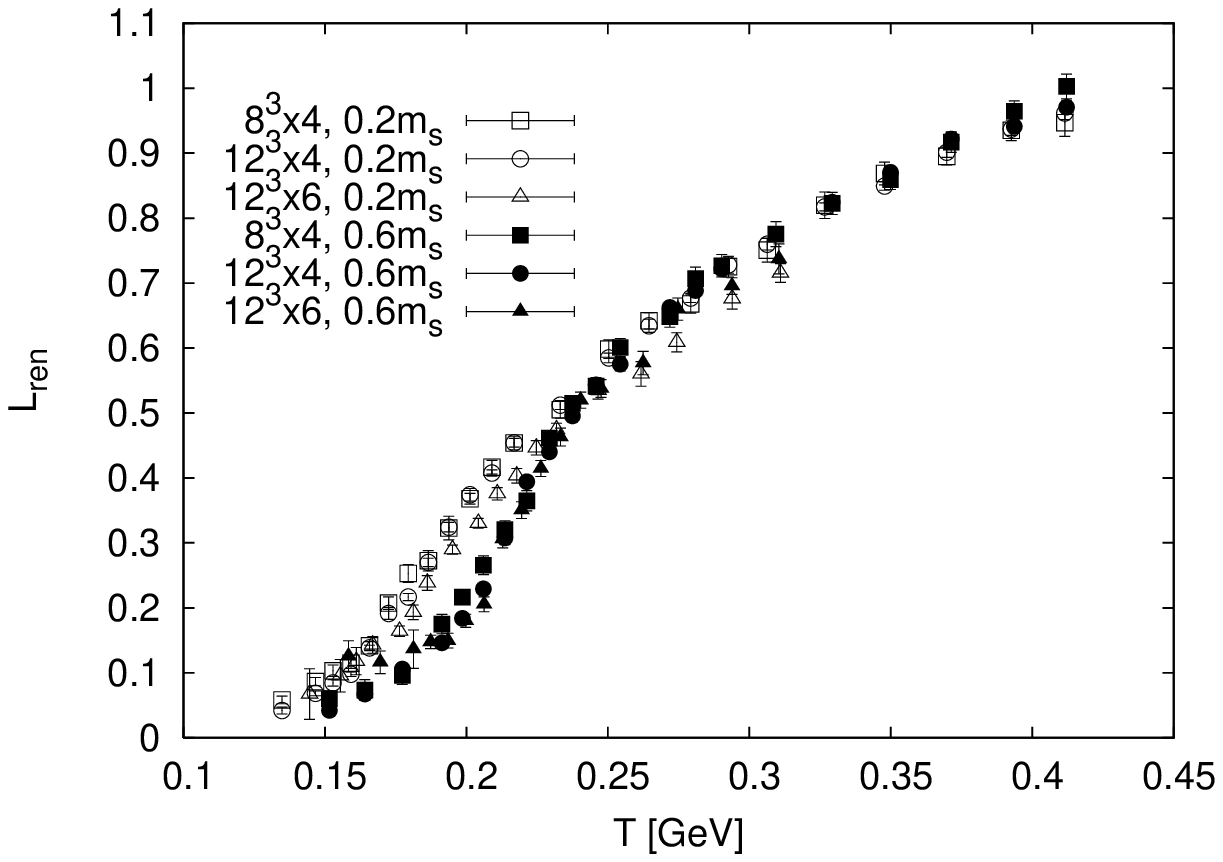}
\caption{The renormalized Polyakov loop calculated on
$12^3 \times 4$ lattices for three different quark masses (left)
and for $m_q=0.2m_s,0.6m_s$ on $8^3 \times 4$, $12^3 \times 4$ and
$12^3 \times 6$ lattices (right). The insertion in the left figure
shows the renormalized Polyakov loop as function of $T/T_c$ (see
text). \label{fig_lren}.}
\end{figure}

The numerical results for $F_{\infty}(T)$ can be easily converted to the
numerical results for the renormalized Polyakov loop $L_{ren}(T)$ which are 
shown in Fig. \ref{fig_lren}. The mass dependence of $L_{ren}(T)$ is the same 
as for $F_{\infty}$, when plotted versus temperature in physical units it
shows mass dependence for $150MeV <T < 250MeV$ and when plotted as function
of $T/T_c$ mass dependence for $T/T_c>1$.
The renormalized Polyakov loop should not depend on the lattices spacing. To check
this we have calculated $L_{ren}(T)$ also on $12^3 \times 6$ and 
$8^3 \times 4$ lattices. The results are summarized in Fig. \ref{fig_lren}, and 
as one can see the $N_{\tau}$ (lattice spacing) dependence of $L_{ren}(T)$ is
quite small.

\section{Conclusions}
In this paper we have presented calculations of the free energy of a static
quark anti-quark pair in three flavor QCD for several
quark masses. We have found that the free
energy gets screened beyond some distance for all temperatures, as 
expected. For small temperature this distance, the effective screening radius,
does not depend on the temperature and is about $0.9fm$. As the temperature
increases the effective screening radius decreases.
Using the asymptotic value of the free energy we have defined 
the renormalized Polyakov loop and proved its scaling with the
lattice spacing. We have also identified the entropy contribution
to the free energy as well as the internal energy at large distances and found that they 
show strong increase at $T_c$. 
We have found substantial quark mass dependence in the vicinity of the
transition.
In the future it will be certainly interesting to
extend this study to larger lattices and to study the dependence of the
entropy contribution as function of the separation which because
of the limited statistics was not possible here. Many properties of the static
quark anti-quark free energies presented in this three flavor study 
turns out to be similar to the preliminary findings in two flavor case
by  the Bielefeld group reported in Ref. \cite{okacz03a}.

\section*{Acknowledgment}
Most of our analysis has been performed using finite temperature
configurations from MILC collaboration. We would like to thank
the members of MILC collaboration for sharing these configurations
as well as for the unpublished data on the zero temperature potential.
Special thanks goes to C. DeTar, U.M. Heller, R. Sugar and D. Touissant for
many helpful communication concerning their results on zero temperature
potential, scale setting and other issues. 
We would like to thank F. Karsch and A. Patk\'os for
careful reading of the manuscript and several valuable suggestions.
This work has been supported by the US Department of Energy under
the contract DE-AC02-98CH10886 and by the SciDAC project.

\end{document}